\documentstyle[12pt]{article}

         \def\la{\lambda}
         \def\t{\tilde}
         \def\be{\begin{equation}}
         \def\bea{\begin{eqnarray}}
         
         \def\o{\over}
         
         \def\r{\rho}

         \def\ee{\end{equation}}
         \def\eea{\end{eqnarray}}
         \def\R{\rm {I\kern-.200em R}}
         \def\C{\rm {I\kern-.520em C}}
         \def\ba{\begin{array}}
         \def\ea{\end{array}}
         \def\be{\begin{equation}}
         \def\bea{\begin{eqnarray}}
         \def\eea{\end{eqnarray}}
         \def\ee{\end{equation}}
         \def\tr{{\rm tr}}

\begin{document}
\begin{titlepage}
%%%%%%%%%%%%%%%%%%%%  The title page   %%%%%%%%%%%%%%%%%%%%%%%%%%%%%%%%%%%%%
\hfill
\vbox{
    \halign{#\hfil         \cr
            hep-th/9707081      \cr
           } % end of \halign
      }  % end of \vbox
\vskip 10 mm
\begin{center}
{ \Large \bf
        Large-$N$ limit of the generalized 2-dimensional}
\vskip 7mm
{ \Large \bf
       Yang-Mills theories}

\vskip 15 mm
{ \bf
M. Alimohammadi$^{a,b,}$\footnote {e-mail:alimohmd@netware2.ipm.ac.ir},
M. Khorrami$^{a,b,c,}$\footnote{e-mail:mamwad@netware2.ipm.ac.ir},
A. Aghamohammadi$^{b,d}$\footnote {e-mail:mohamadi@netware2.ipm.ac.ir}}
\vskip 10 mm
{\it

 $^a$ Department of Physics, Tehran University,
             North-Kargar Ave.\\
             Tehran, Iran \\
 $^b$ Institute for Studies in Theoretical Physics and Mathematics,\\
           P.O.Box  5531, Tehran 19395, Iran\\
 $^c$ Institute for Advanced Studies in Basic Sciences,
 P.O.Box 159,\\
Gava Zang, Zanjan 45195, Iran\\
 $^d$ Department of Physics, Alzahra University,  Tehran 19834, Iran  }
\end{center}
\vskip 3cm
\begin{abstract}
Using the standard saddle-point method, we find an explicit
relation for the large-$N$ limit of the free energy of an arbitrary generalized
2D Yang-Mills theory in the weak ($A<A_c$) region. In the strong ($A>A_c$)
region, we investigate carefully the specific fourth Casimir theory, and
show that the ordinary integral equation of the density function is not adequate
to find the solution. There exist, however, another equation which restricts
the parameters. So one can find the free energy in strong region and show
that the theory has a third order phase transition.

\end{abstract}

\end{titlepage}
\newpage
%%%%%%%%%%%%%%%%%%%  The body of the paper            %%%%%%%%%%%%%%%%%%%%%

\section{\bf Introduction}
The pure 2D Yang-Mills ($\tr(F^2)$) theory (YM$_2$) defined on compact Riemann
surfaces, is characterized by its invariance under area preserving diffeomorphism
and the fact that there are no propagating degrees of freedom.
This theory is not unique in this sense, and it is possible to generalize it
without losing these properties.

In an equivalent formulation of YM$_2$ theory, one can use $i\tr(BF)+\tr(B^2)$ as
the Lagrangian, where $B$ is an auxiliary pseudo-scalar field in the adjoint
representation of the gauge group. Path integration over the field $B$ leaves
an effective Lagrangian of the form $\tr(F^2)$ [1]. Now, a generalized 2D
Yang-Mills theory (gYM$_2$) is a theory with above Lagrangian, in which the
term $\tr(B^2)$ is replaced by an arbitrary class function $\Lambda (B)$ .

In [2], where these theories were first introduced, the partition function
of gYM$_2$ have been obtained by considering its action as a perturbation of
the topological theory at zero area. In [3], these theories have been
coupled to fermions, to obtain the generalized QCD$_2$. Generalizing the
Migdal's suggestion about the local factor of plaquettes [4], the authors of
[5] have found the partition function and the expectation values of the Wilson
loops for gYM$_2$'s. In [6], these quantities and also the generating
functional of the field strength have been calculated for arbitrary two
dimensional orientable and nonorientable surfaces, using the standard path
integral method.

One of the important features of YM$_2$, and also gYM$_2$'s, is their behaviour
in the case of large gauge groups, e.g., the large $N$ behaviour of $SU(N)$
(or $U(N)$) gauge theories. On one hand, it is interesting to study the
relation between these theories at large $N$ and the string theory. Such
investigations began in [7] and [8] for YM$_2$ and in [5] for gYM$_2$. In these
papers, it was shown that the coefficients of $1/N$ expansion of the partition
function of $SU(N)$ gauge theories are determined by a sum over maps from a
two-dimensional surface onto the two-dimensional target space.

Another thing of interest, also related to the above mentioned one, is to
study the large $N$ behaviour of the free energy of these theories. This is done
by replacing the sum over the irreducible representations of $SU(N)$ (or $U(N)$),
appearing in the expressions of partition function and Green functions, by a
path integral over continuous Young tableaux, and calculating the
area-dependence of the physical quantities from the saddle-point configuration.
In [9], It was shown that the free energy of $U(N)$ YM$_2$ on a sphere with
surface area $A<A_c=\pi^2$ has a logarithmic behaviour. Later, the authors of
[10] calculated the free energy for surface area $A>\pi^2$, and showed that the
YM$_2$ has a third order phase transition at the critical surface area. Such
kind of calculation for gYM$_2$ began in [11], and the authors claimed that
these theories have a rich phase structure. They also discussed the general
behaviour of the density function $\rho$ for $A<A_c$,
and brought some comments for the $A>A_c$ region.

In this paper, we first complete the calculation of the free energy of $U(N)$
gYM$_2$ for $A<A_c$ region for an arbitrary gYM$_2$ on sphere and derive an
exact solution for the density function. But the main part of the paper is about
the $A>A_c$ region of a specific fourth Casimir model. It is seen that here,
unlike the YM$_2$ model, the density function $\rho$ has two maxima, for
$A<A_c$. So some new features arise in the phase transition, which must be
handled more carefully. At the end we show, however, that for this particular
model there exist a third order phase transition, just as in the case of
YM$_2$. This is another similarity between these two theories.

\section {\bf Large-$N$ behaviour of gYM$_2$ at $A<A_c$ }

The partition function of the gYM$_2$ on a sphere is [5,6]
\be Z=\sum_r d_r^2e^{-A\Lambda (r)}, \ee
where $r$'s label the irreducible representations of the gauge group, $d_r$
is the dimension of the $r$'th representation, $A$ is the area of the sphere
and $\Lambda (r) $ is :
\be \Lambda (r) = \sum_{k=1}^p {a_k \over {N^{k-1}}}C_k(r),\ee
in which $C_k$ is the $k$'th Casimir of group, and $a_k$'s are arbitrary
constants. Now consider the gauge group
$U(N)$ and parametrize its representation by $n_1\geq n_2\geq\cdot \cdot
\cdot\geq n_N$, where $n_i$ is the length of the $i$'th row of the Young
tableau. It is found that [12]
$$d_r=\prod_{1\leq i < j \leq N } (1+{{n_i-n_j}\over {j-i}}) $$
\be C_k=\sum_{i=1}^N[(n_i+N-i)^k-(N-i)^k].\ee
To make the partition function (1) convergent, it is necessary that $p$ in
eq.(2) be even and $a_p >0$.

Now, following [9], we can write the partition function (1) at large $N$, as
a path integral over continuous parameters. We introduce the continuous
function :
\be \phi (x)=-n(x)-1+x, \ee
where
\be 0 \leq x:=i/N \leq 1 \ \ \ \ {\rm and }  \ \ \ \ n(x):=n_i/N. \ee
The partition function (1) then becomes
\be Z=\int \prod_{0\leq x \leq 1} d\phi(x)e^{S[\phi(x)]},\ee
where
\be S(\phi)=N^2\{-A\int^1_0dxG[\phi(x)]+\int_0^1dx\int_0^1dy \ \ {\rm log}|\phi (x)-\phi (y)|
\ \ \},\ee
apart from an unimportant constant, and :
\be G(\phi )=\sum_{k=1}^p(-1)^ka_k\phi^k.\ee
As $N\rightarrow \infty $, the partition function (7) is determined by the
configuration which maximizes $S$. The saddle point equation for $S$ is :
\be g[\phi (x)]={\rm P}\int_0^1{dy \over {\phi (x) - \phi (y)}},\ee
where:
\be g(\phi )={A\over 2} G'(\phi ), \ee
and P indicates the principal value of the integral. Introducing the density
\be \rho [\phi (x)]={dx \over d\phi (x)},\ee
the eq.(9) is reduced to:
\be g(z)={\rm P}\int^a_b {\rho (\la )d \la \over z-\la},\ee
along with the normalization condition
\be\int^a_b\rho (\la )d \la=1 .\ee
The condition $n_1\geq n_2\geq\cdot \cdot \cdot\geq n_N$ imposes the following
condition on the density $\rho (\la )$ :
\be \rho (\la ) \leq 1 . \ee
To solve eq.(12), we define the function $H(z)$ in complex $z$-plane [13]
,
\be H(z):=\int_b^a {\rho (\la )d \la\over z-\la}.\ee
This function is analytic on the complex plane exept for a cut at $[b,a]$. There
, one has
\be H(z\pm i \epsilon )=g(z)\mp i\pi \rho (z) \ \ \ \  \  \ b\leq z\leq a.\ee
$H$ is found to be [10,14] :
\be H(z)={1\over 2\pi i } \sqrt{(z-a)(z-b)} \oint_c {g(\la ) d \la \over (z-\la )
\sqrt{(\la -a )(\la -b)}},\ee
where $c$ is a contour encircling the cut $[b,a]$, and excluding $z$ and $g(\la )$
is defined through eq.(10). Deforming
$c$ to a contour around the point $z$ and the contour $c_\infty$ (a contour at
the infinity ), one finds:
\be H(z)=g(z)-\sqrt{(z-a)(z-b)}\sum_{m,n,q=0}^\infty {(2n-1)!!(2q-1)!!\over
2^{n+q}n!q!(n+q+m+1)!}a^nb^qz^mg^{(n+q+m+1)}(0),\ee
where $g^{(n)}$ is the $n$-th derivative of $g$. It follows from (15) and (13)
that $H(z)$ behaves like $1/z$ at $z\rightarrow \infty$, or
$(z-a)^{-1/2}(z-b)^{-1/2}H(z)$ behaves like $1/z^2$ at $z\rightarrow \infty$.
Using this, one can expand $(z-a)^{-1/2}(z-b)^{-1/2}H(z)$, equate the
coefficients of $1/z$ and $1/z^2$ equal to $0$ and $1$, respectively, and
arrive at
\be \sum_{n,q=0}^\infty {(2n-1)!!(2q-1)!!\over 2^{n+q}n!q!(n+q)!}
    a^nb^qg^{(n+q)}(0)=0,\ee
and
\be \sum_{n,q=0}^\infty {(2n-1)!!(2q-1)!!\over 2^{n+q}n!q!(n+q-1)!}
    a^nb^qg^{(n+q-1)}(0)=1.\ee
These equations determine $a$ and $b$. From (16), $\rho$ is determined :
\be \rho (z)={\sqrt{(a-z)(z-b)}\over \pi}\sum_{m,n,q=0}^\infty
{(2n-1)!!(2q-1)!!\over 2^{n+q}n!q!(m+n+q+1)!}a^nb^qz^mg^{(m+n+q+1)}(0). \ee
Defining the free energy as
\be F:=-{1\over N^2}{\rm ln}Z,\ee
one has
\be F'(A)=\int_0^1dx \ \ G[\phi (x)]=\int_b^ad\la \ \ G(\la)\rho (\la) . \ee
Expanding $H(z)$ for $z\rightarrow \infty $, one can find the integrals
\be \int_b^ad\la \ \ \la^n\rho(\la), \ee
by using the eqs.(15) and (18). Knowing these, one can calculate $F'(A)$ through (23).

It is also worth mentioning that if $G$ is an even function, then $b=-a$,
$\rho$ is an even function, and equations (17), (18), (20) and (21) become
$$ H(z)={1\over 2\pi i } \sqrt{z^2-a^2} \oint_c {g(\la ) d \la \over (z-\la )
\sqrt{\la^2 -a^2}}, \eqno (17)' $$
$$ H(z)=g(z)-\sqrt{z^2-a^2}\sum_{n,q=0}^\infty {(2n-1)!!\over
2^nn!(2n+q+1)!}a^{2n}z^qg^{(2n+q+1)}(0),\eqno (18)'$$
$$ \sum_{n=0}^\infty {(2n-1)!!\over 2^nn!(2n-1)!}
    a^{2n}g^{(2n-1)}(0)=1,\eqno (20)' $$
and
$$ \rho (z)={\sqrt{a^2-z^2}\over \pi}\sum_{n,q=0}^\infty
{(2n-1)!!\over 2^nn!(2n+q+1)!}a^{2n}z^qg^{(2n+q+1)}(0). \eqno (21)'$$
As the simplest example of gYM$_2$, consider $G(\phi )=\phi^4$. Using the
above relations, one finds
$$ \rho (z)={2A \over \pi}({a^2 \over 2}+z^2)\sqrt{a^2-z^2}, $$
\be a=({4 \over 3A})^{1/4}, \ee
$$ F'(A)={1 \over 4A}. $$
The density $\rho $ has a minimum at $z=0$ and two maxima at $z_0^{(\pm)}=\pm
a/ \sqrt{2}$. Now as $\rho(z_0^{(\pm)})=\sqrt{2}a^3A/\pi$, if $A>A_c=27\pi^4/256$
, then the condition $\rho \leq 1$ is violated. So the solution (25) is valid
only in the  region $A\leq A_c$.

It is worth noting that the function $F'(A)$ for $G(\phi )=\phi^k$ can be simply
found by rescaling the field $\phi$ by $A^{1/k}$ in eq.(7) (as noted in [10] )
, after which the only $A$-dependent term of $S$ is log$A^{1/k}$ and therefore
$F'(A)={1 \over kA}$.

\section {\bf The $G(\phi )=\phi^4$ model in the region $A>A_c$ }

In this region, the solution (25) is not correct any more. As $\r (z)$ for
$A<A_c$ has two symmetric maxima, we use the following ansatz for $\r $ :
\be
\r_s(z)=\cases{ 1,&$z\in [-b,-c]\bigcup
[c,b]=:L'$  \cr
\tilde{\r}_s(z),&$ z\in [-a,-b]\bigcup [-c,c] \bigcup [b,a]=:L.$\cr
}
\ee
Putting this in (7), we have
\be S(\r_s)=N^2 \left[ \ \ -A\int_{-a}^adz \ \ \r_s(z)z^4+\int_{-a}^adz
\int_{-a}^adw\r_s(z)\r_s(w) {\rm log}|z-w| \ \ \right]. \ee
To maximize this along with the condition (13), we introduce the Lagrange
multiplier $\mu $ and the functional $\tilde{S}$ :
\be \tilde{S}=S+N^2\mu\left[ \ \ \int_{-a}^adz \  \ \r_s(z)-1  \ \ \right]. \ee
The variation of $\tilde{S}$ is
\be \delta \tilde{S}=N^2\int_{-a}^adz \left[ \ \ -Az^4+2\int_{-a}^adw \ \ \r_s(w)
{\rm log} |z-w|+\mu \ \ \right] \delta\r_s +N^2\left[ \ \ \int_{-a}^adz \ \
\r_s(z)-1 \ \ \right]\delta \mu, \ee
where
\be
\delta\r_s(z)=\cases{ 0,&$ z\in L'$ \cr
\delta \tilde{\r}_s(z), &$ z\in L.$\cr}\ee
Equating $\delta \tilde{S}$ to zero, we have
\be {A\o 2}z^4-\int_{-a}^adw \ \ \r_s(w) {\rm log}|z-w|={\mu\o 2}, \ \ \ \ \
\ \ z\in L , \ee
and
\be \int_{-a}^adz \ \ \r_s(z)=1. \ee
Differentiating (31) with respect to $z$, we have
\be 2Az^3={\rm P}\int_{-a}^adw \ \ {\r_s(w) \o z-w}, \ \ \ \ \ \ z\in L. \ee
This equation is, however, not equivalent to (31), since $L$ is a disconnected
set. In fact, (31) is equivalent to (33) and the following equation
\be {A\o 2}(b^4-c^4)-\int_{-a}^adw \ \ \r_s(w) {\rm log}|{b-w \o c-w}|=0 , \ee
which is difference of eq.(31) at $z=b$ and $z=c$. The
above equation can also be written as
\be \int_c^bdz \left[ \ \ 2Az^3-{\rm P} \int_{-a}^adw \ \ {\r_s(w)\o z-w}
\right]=0. \ee
So our task is to solve the eq.(33), along with the conditions (32) and (35). To do
so, we use the procedure of the previous section:
$$ H_s(z):={\rm P}\int_{-a}^adw \ \ {\r_s(w)\o z-w}=
\int_Ldw \ \ {\tilde{\r}_s(w)\o z-w}+{\rm log}{z+b\o z+c}+{\rm log}{z-c\o z-b}$$
\be := \tilde{H}_s(z)+{\rm log}{z+b\o z+c}+{\rm log}{z-c\o z-b}\ee
where $\tilde{H}_s$ has a three cut singularity at $z\in L$. Similar to eq. (17), the
solution of $\t{H}_s$ is [14] :
\be \t{H}_s(z)={1\o 2\pi i} \sqrt{(z^2-a^2)(z^2-b^2)(z^2-c^2)}\oint_{c_L}
{\t{g}(\la)d\la\o (z-\la)\sqrt{(\la^2-a^2)(\la^2-b^2)(\la^2-c^2)}}, \ee
where
$$ \t{g}(z):=g(z)-{\rm log}{z+b\o z+c}-{\rm log}{z-c\o z-b} $$
\be =2Az^3-{\rm log}{z+b\o z+c}-{\rm log}{z-c\o z-b}, \ee
and $c_L$ is a contour encircling the three distinct intervals of $L$. Deforming
$c_L$ to a contour at infinity and three contours encircling the point $z$ and
the two intervals $[-b,-c]$ and $[c,b]$, respectively, one can determine
$\t{H}_s$ to be
\be \t{H}_s(z)=\t{g}(z)-2\sqrt{(z^2-a^2)(z^2-b^2)(z^2-c^2)}\left(A+\int_c^b
{\la d\la\o (z^2-\la^2)\sqrt{(a^2-\la^2)(b^2-\la^2)(\la^2-c^2)}} \right), \ee
and
\be H_s(z)=2Az^3-2\sqrt{(z^2-a^2)(z^2-b^2)(z^2-c^2)}\left(A+\int_c^b
{\la d\la\o (z^2-\la^2)\sqrt{(a^2-\la^2)(b^2-\la^2)(\la^2-c^2)}} \right). \ee
Using the definition of $H_s(z)$ in (36), we arrive at the following expansion
for $H_s(z)$ at large $z$
\be H_s(z)={1\o z}+{1\o z^3} \int_{-a}^a\r_s(\la)\la^2d\la+{1\o z^5}
F_s'(A)+\cdot\cdot\cdot, \ee
where $F_s$ is the free energy in the strong region. Therefore, one can
expand \linebreak $H_s(z)/\sqrt{(z^2-a^2)(z^2-b^2)(z^2-c^2)}$ and demand that it behaves
like $1/z^4$ at large $z$. This gives
\be A(a^2+b^2+c^2)=
2\int_c^b{\la d\la\o \sqrt{(a^2-\la^2)(b^2-\la^2)(\la^2-c^2)}}, \ee
and
\be A\left[ \ \ {3\o 4}(a^4+b^4+c^4)+{1\o 2}(a^2b^2+b^2c^2+c^2a^2) \ \ \right]
-2\int_c^b{\la^3 d\la\o \sqrt{(a^2-\la^2)(b^2-\la^2)(\la^2-c^2)}}=1. \ee
Finally, the $1/z^5$ coefficient of $H_s(z)$ is $F_s'(A)$. So,
$$ F_s'(A)={A\o 16}\Big[ \ \ {5\o 4}(a^8+b^8+c^8)-{1\o 2}(a^4b^4+a^4c^4+b^4c^4)
+(a^2b^2c^4+a^2c^2b^4+b^2c^2a^4) $$
$$ -(a^2b^6+a^2c^6+b^2a^6+b^2c^6+c^2a^6+c^2b^6) \ \ \Big] $$
$$ +{1\o 8}\left[ \ \ a^6+b^6+c^6-(a^2b^4+a^2c^4+b^2a^4+b^2c^4+c^2a^4+c^2b^4)
 +2a^2b^2c^2 \ \ \right]\int_c^b{\la d\la\o R(\la)} $$
\be +{1\o 4}\left[ \ \ a^4+b^4+c^4-2(a^2b^2+a^2c^2+b^2c^2)
 \ \ \right]\int_c^b{\la^3 d\la\o R(\la)} +(a^2+b^2+c^2)
\int_c^b{\la^5 d\la\o R(\la)}
-2\int_c^b{\la^7 d\la\o R(\la)}, \ee
where
\be R(\la ):=\sqrt{(a^2-\la^2)(b^2-\la^2)(\la^2-c^2)}. \ee
Note that (42) and (43) are just two equations for three unknowns $a$, $b$ and
$c$. This is the same situation encountered in [15]. There, the authors have
considered an asymmetric ansatz for the density of YM$_2$, and have found three
equations for four parameters. We will discuss this problem in [16]. Now, for
the case in hand, the third equation is (35), which using eqs. (36) and (37)
can be rewritten as
\be A\int_c^b R(z)dz+\int_c^b dz \ \ {\rm P}\int_c^b \ \
  {R(z)\la d\la \o (z^2-\la^2)R(\la)}=0. \ee
Now we have three equations ( (42),(43), and (46) ) for three unknowm parameters
$a$, $b$ and $c$.

To study the structure of the phase transition, we use the following change
of variables
$$ c=s(1-y), $$
\be b=s(1+y), \ee
$$ a=s \sqrt{2+e}. $$
Here, $e$ and $y$ are equal to zero at the transition point ($A=A_c$).
Now, one can expand the equations (42), (43) and (46) and obtain
\be 4As^2-{\pi \o s} +(As^2+{\pi \o 2s})e-{3\pi e^2 \o 8s}+(2As^2-{5\pi \o 4s})
y^2+{21\pi ey^2 \o 8s}-{249\o 64}{\pi y^4\o s}=0 ,\ee
$$ 7As^4-1-\pi s+(4As^4+{\pi s \o 2})e+({3As^4 \o 4}-{3\pi s\o 8})e^2+(10As^4
-{13 \pi s \o 4})y^2 $$
\be +(As^4+{37\pi s \o 8})ey^2+(2As^4-{545 \o 64}\pi s)y^4=0 , \ee
and
$$ 2\pi-8As^3-4Aes^3-2\pi e+Ae^2s^3+2\pi e^2+{1\o 4}(16As^3+17\pi-8Aes^3+46\pi
e)y^2 $$
\be +(2As^3+{491\o 32}\pi )y^4=0,\ee
respectively. These expansions are up to order $y^4$, or $e^2$ (as we will see
$e$ is of the order $y^2$). Solving eq. (48) for $s$ results
\be s=({\pi \o 4A})^{1/3}\left( 1-{e\o 4}+{e^2\o 8}+{y^2\o 4}-{11ey^2\o 16}
+{71y^4\o 64} \right), \ee
from which, one obtains ( using eq.(50) )
\be e={5\o 2}y^2-{3\o 16 }y^4. \ee
This shows that $e$ is, in fact, of the order $y^2$. Substituting these into
the eq.(49), we arrive at
\be y^2={\alpha\o 3}+{7\o 96}\alpha^2, \ee
and
\be e={5\o 6}\alpha +{31\o 192}\alpha^2, \ee
where $\alpha $ is the reduced area, $\alpha ={A-A_c\o A_c}$. Now $F_s'(A)$
in eq. (44) can be calculated. The result is
\be F_s'(A)={1\o 4A}(1+{4\o 27}\alpha^2 +\cdot\cdot\cdot ). \ee
Comparing this with $F_w'(A)$, eq.(25), it is seen that
\be F_s'(A)-F_w'(A)={1\o 27A_c}({A-A_c\o A_c})^2+\cdot\cdot\cdot . \ee
This shows that we have a third order phase transition, which is the same as
the ordinary YM$_2$ [10].
\vskip 1cm
{\bf Acknowledgement} M. Alimohammadi would like to thank the research vice-chancellor
of Tehran University, this work was partially supported by them.


\vskip 1cm

\begin{thebibliography}{99}

\bibitem {1} M. Blau and G. Thompson, ``Lectures on $2d$ Gauge Theories'',
             Proceedings of the 1993 Trieste Summer School on High Energy
             Physics and Cosmology, World Scientific,
             Singapore (1994) 175-244.
\bibitem {2} E. Witten, J. Geom. Phys. {\bf 9} (1992) 303.
\bibitem {3} M. R. Douglas, K. Li and M. Staudacher, Nucl. Phys. {\bf B240}
             (1994) 140.
\bibitem {4} A. Migdal, Zh. Eskp. Teor. Fiz. {\bf 69} (1975) 810
             (Sov. Phys. JETP {\bf 42}, 413).
\bibitem {5} O. Ganor, J. Sonnenschein, and S. Yankielowicz,
             Nucl. Phys. {\bf B434} (1995) 139.
\bibitem {6} M. Khorrami and M. Alimohammadi, Mod. Phys. Lett. {\bf A12}
             (1997) 2265.
\bibitem {7} D. J. Gross, Nucl. Phys. {\bf B400} (1993) 161.
\bibitem {8} D. J. Gross and W. Taylor, Nucl. Phys. {\bf B400} (1993) 181.
\bibitem {9} B. Rusakov, Phys. Lett. {\bf B303} (1993) 95.
\bibitem {10} M. R. Douglas and V. A. Kazakov, Phys. Lett. {\bf B319} (1993)
             219.
\bibitem {11} B. Rusakov and S. Yankielowicz, Phys. Lett. {\bf B339} (1994)
             258.
\bibitem {12} V. S. Popov and A. M. Perelomov, Soviet Math. Dokl. {\bf 8}
            (1967) 712.
\bibitem {13} E. Brezin, C. Itzykson, G. Parisi and J. B. Zuber, Commun.
             Math. Phys. {\bf 59} (1978) 35.
\bibitem {14} D. Gakhov, ``Boundary problems'', Russian edition
             (Nauka, 1975);\\
              A. C. Pipkin, ``A course on integral equations''
             (Springer, Berlin, 1991).
\bibitem {15} J. A. Minahan and A. P. Polychronakos, Nucl. Phys. {\bf B422}
             (1994) 172.
\bibitem {16} A. Aghamohammadi, M. Alimohammadi and M. Khorrami,
             ``Uniqueness of the minimum of the free energy of the 2D
             Yang-Mills theory at large N'', hep-th/9707080
\end{thebibliography}
\end{document}